\documentclass[12pt]{article}

\usepackage{epsf}
\usepackage{amsmath}
\usepackage{graphics}
\usepackage{amsfonts}
\usepackage{amssymb}
\usepackage{latexsym}
\usepackage{color}
\input{colordvi.tex}

\renewcommand{\thefootnote}{\#\arabic{footnote}}

\begin{document}
\setcounter{footnote}{0}

\begin{titlepage}

\begin{center}


\vskip .5in

{\Large \bf
Exact gravitational lensing by cosmic strings with junctions
}

\vskip .45in

{\large
Teruaki Suyama
}

\vskip .45in

{\em
Institute for Cosmic Ray Research, 
University of Tokyo, Kashiwa 277-8582, Japan
}

\end{center}

\vskip .4in

\begin{abstract}
We point out that the results by Brandenberger et al. in 
Phys.Rev.D77:083502(2008) that
the geometry around the straight cosmic strings
with stationary junctions is flat to linear order in the
string tension can be immediately extended to any order.
\end{abstract}
\end{titlepage}

\renewcommand{\thepage}{\arabic{page}}
\setcounter{page}{1}
\renewcommand{\thefootnote}{\#\arabic{footnote}}

\section{Introduction}
Cosmic strings may be produced in the early universe.
If these strings are fundamental strings,
F-strings,
or D1-branes,
D-strings,
which are generic end products of the brane inflation,
these strings can combine to form bound states of p F-strings
and q D-strings,
(p,q)-strings \cite{Dvali:2003zj, Copeland:2003bj, Leblond:2004uc, Firouzjahi:2006vp, Dasgupta:2007ds}.
When two (p,q)-strings intersect,
a junction is formed \cite{Jackson:2004zg,Copeland:2007nv,Bevis:2008hg,Rajantie:2007hp}.

In \cite{Shlaer:2005ry},
the authors gave a simple geometrical prescription to obtain the location
of images of the source lensed by the strings with a junction.
In \cite{Brandenberger:2007ae},
the explicit form of the metric around the strings with stationary
junctions was provided within the linear perturbation.
It was also shown that,
like in the case of an infinite straight string \cite{Vilenkin:1981zs, Gott:1984ef}, 
the space-time is locally flat and has deficit angles around the
strings determined by the string tensions.

In this short note,
we point out that the flatness of the geometry can be shown without
invoking a perturbative expansion.

\section{Flatness of the geometry}

We start from the action of the strings with junctions
\begin{eqnarray}
S=-\sum_{A=1}^N \mu_A \int d^2\sigma_A \sqrt{-\gamma_A}. \label{action}
\end{eqnarray}
Here $N$ is the total number of the strings,
$\sigma_A^a (a=0,1)$ is the two-dimensional coordinate on 
the $A$-th string worldsheet and 
\begin{eqnarray}
\gamma_{A ab} \equiv g_{\mu \nu}(X_A) \frac{\partial X^\mu_A}{\partial \sigma_A^a} \frac{\partial X^\nu_A}{\partial \sigma_A^b}, \label{induced-metric}
\end{eqnarray}
is the induced metric on the $A$-th string.
$\mu_A$ is the tension of the $A$-th string.
The strings are assumed to be infinitely thin.
Taking the variation of $S$ with respect to $g_{\mu \nu}$ yields
the energy-momentum tensor of the strings,
which is given by
\begin{eqnarray}
T^{\mu \nu}(x)=-\sum_{A=1}^N \frac{\mu_A}{\sqrt{-g(x)}} \int d^2\sigma_A \sqrt{-\gamma_A} \gamma_A^{ab} \frac{\partial X^\mu_A}{\partial \sigma_A^a} \frac{\partial X^\nu_A}{\partial \sigma_A^b} ~\delta^{(4)} (x-X_A ). \label{energy-tensor}
\end{eqnarray}

We are interested in the metric around the static strings with
junctions.
For the junction to be static,
the forces exerted on the junction from attached
strings must balance,
which we will assume hereafter.
Derivation of the force balance condition will be given later.
The static condition implies that there exists a coordinate system 
in which $g_{0i}=0$ and the other metric components do not depend on $t$.
As the metric around the strings,
we use the following ansatz:
\begin{eqnarray}
ds^2=-dt^2+h_{ij} ({\vec x}) dx^i dx^j. \label{metric}
\end{eqnarray}
$g_{00}$ is not perturbed by the strings,
which was explicitly shown at the linearized level \cite{Brandenberger:2007ae}.
Here we assume that this holds in fully non-linear treatment,
which will turn out to be consistent with basic equations.

Let us choose the time coordinate on each worldsheet as $\sigma_A^0=t$,
i.e. $X^0_A (\sigma_A)=\sigma_A^0$.
Then the static condition means that $X^i(\sigma_A)$ is independent of 
$\sigma_A^0$.
As for the spatial coordinate,
we take $\sigma_A^1$ so that it represents the length of the corresponding
string,
which gives ${(d\sigma_A^1)}^2=h_{ij}dX_A^i dX_A^j$.
In these worldsheet coordinates,
the induced metrics are given by
\begin{eqnarray}
\gamma_{A00}=-1,~~~\gamma_{A01}=0,~~~\gamma_{A11}=1. \label{induced-metric2}
\end{eqnarray}

Let us first consider the Einstein equations, 
\begin{eqnarray}
R_{\mu \nu}=8\pi G S_{\mu \nu}\equiv 8\pi G \left( T_{\mu \nu}-\frac{1}{2} g_{\mu \nu}T\right). \label{einstein-eq1}
\end{eqnarray}
Substituting Eqs.~(\ref{metric}) and (\ref{induced-metric2}) into
Eq.~(\ref{energy-tensor}) yields $S_{0\mu}=0$.
From the metric (\ref{metric}),
we find $R_{0\mu}=0$ for arbitrary $h_{ij}({\vec x})$.
Hence $0-\mu$ components of the Einstein equations are automatically
satisfied.
Meanwhile,
$i-j$ components of the Einstein equations give equations for $h_{ij}({\vec x})$,
\begin{eqnarray}
R^{(3)}_{ij}=8\pi G S_{ij}, \label{einstein-eq2}
\end{eqnarray}
where $R^{(3)}_{ij}$ is the Ricci tensor calculated from the three-dimensional
metric $h_{ij}({\vec x})$.
At the linear order,
expanding the spatial metric as $h_{ij}=\delta_{ij}+\delta h_{ij}$,
the solution of the equation (\ref{einstein-eq2}) can be written as
$\delta h_{ij}({\vec x})=-4G \int d^3 y S_{ij}({\vec y})/|{\vec x}-{\vec y}|$,
which agrees with the results in \cite{Brandenberger:2007ae}.
Solving the equations (\ref{einstein-eq2}) for $h_{ij}({\vec x})$ beyond
the linear approximation is cumbersome.
However,
the knowledge of the explicit expression for $h_{ij}({\vec x})$ is
not necesarry for our argument and we skip solving Eq.~(\ref{einstein-eq2}).

The Riemann tensor $R_{\mu \nu \lambda \rho}$ for the metric 
(\ref{metric}) is given by $R_{\mu \nu \lambda 0}=0,~R_{ijk\ell}=R^{(3)}_{ijk\ell}$,
where $R^{(3)}_{ijk\ell}$ is the Riemann tensor for $h_{ij}$.
In three-dimensional space,
the Riemann tensor is completely written by the Ricci tensor,
\begin{eqnarray}
R^{(3)}_{ijk\ell}=-2 \left( h_{i[ \ell} {R^{(3)}}_{k]j}+h_{j[k}{R^{(3)}}_{\ell] i}-R^{(3)} h_{i[k} h_{\ell]j} \right),
\end{eqnarray}
where $[\cdots ]$ means the anti-symmetrization with respect to 
the indices in the bracket.
Outside the strings,
$R^{(3)}_{ij}$ vanishes from the Einstein equations (\ref{einstein-eq2}),
which gives $R^{(3)}_{ijk\ell}=0$.
Therefore the space-time is flat around the strings.

We have implicitly assumed that the strings are straight and
force balanced.
Here we show that these assumptions are validated from an equation of 
motion for the strings and a boundary condition at the junction.
Variation of Eq.~(\ref{action}) with respect to $X^\mu_A$ yields,
\begin{eqnarray}
\frac{\partial}{\partial \sigma^a_A} \left( \sqrt{-\gamma_A} \gamma^{ab}_A \frac{\partial X^\mu_A}{\partial \sigma^b_A} \right)+\sqrt{-\gamma_A} \gamma^{ab}_A \Gamma^\mu_{~\lambda \rho} \frac{\partial X^\lambda_A}{\partial \sigma^a_A} \frac{\partial X^\rho_A}{\partial \sigma^b_A}=0, \label{string-eom}
\end{eqnarray}
with the boundary condition
\begin{eqnarray}
\sum_{A'} \mu_{A'} \sqrt{-\gamma_{A'}}~ n^a_{A'} \frac{\partial X^\mu_{A'}}{\partial \sigma^a_{A'}} \bigg |_{\rm junction}=0, \label{boundary}
\end{eqnarray}
for each junction \footnote{The boundary condition can be derived as follows.
For simplicity,
we assume that there is only one junction in the string network (Extension
to the case of multi-junctions is trivial).
We first attach a point mass of its mass $m$ to the junction.
Denoting the coordinate of the point mass as $Z^\mu $,
we have to add the terms
\begin{eqnarray}
-m\int_{\rm junc} d\tau-\sum_A \int_{\rm junc}d\tau ~\lambda_{A \mu} \left( X^\mu_A (\sigma_A(\tau))-Z^\mu (\tau) \right), \nonumber
\end{eqnarray}
to Eq.~(\ref{action}),
where $\tau$ is the proper time and $\sigma_A(\tau)$ denotes the world-sheet coordinate of the junction.
$X^\mu_A(\sigma_A(\tau))=Z^\mu(\tau)$ are the constraints that
the strings are attached to the point mass.
$\lambda_{A\mu}$ are the Lagrange multipliers.
Because no mass point is assumed to be attached to the junction in the main text,
we have to take the limit $m \to 0$.
By taking the limit $m \to 0$ in the equation of motion for $Z^\mu$,
we obtain a relation $\sum_A \lambda_{A \mu}=0$.
Substitution of this relation into the boundary conditions
which are derived from the variation of $X^\mu_A$ yields Eq.~(\ref{boundary}).
}.
Here $n^a_A$ is the unit vector normal to the world-line of the junction and
$A'$ means that we take the sum only for strings attached to the junction
we are considering.
Imposing the static condition,
the equation (\ref{string-eom}) becomes
\begin{eqnarray}
\frac{d^2 X^i_A (\sigma^1_A)}{d \sigma^1_A d\sigma^1_A}+\Gamma^i_{~jk} \frac{dX^j_A (\sigma^1_A)}{d\sigma^1_A}\frac{dX^k_A (\sigma^1_A)}{d\sigma^1_A}=0, \label{string-eom2}
\end{eqnarray}
which is the geodesic equation in the three dimensional space.
Because the space is locally flat,
the straight configuration of the strings is a solution of the equations.
The boundary condition becomes
\begin{eqnarray}
\sum_{A'} \mu_{A'} \frac{dX^i_{A'}}{d\sigma^1_{A'}}\bigg |_{\rm junction}=0, \label{boundary2}
\end{eqnarray}
which is exactly the force balance condition.
To maintain the static condition,
the force balance condition must be satisfied.
Note that the force balance condition can be also derived from the conservation law
$\nabla_\mu T^{\mu \nu}=0$.

The space-time is not globally flat and has conical structure.
The deficit angle $\Delta_A$ around the A-th string can be written 
as $\Delta_A = \frac{1}{2} \int_{\rm plane} d^2x \sqrt{g^{(2)}}R^{(2)}$,
where the metric and the Ricci scalar in the integral are defined
on the two-dimensional plane perpendicular to the string.
After the use of the Einstein equations to replace $R^{(2)}$ 
with the energy-momentum tensor of the string,
the well-known result $\Delta_A =8\pi G\mu_A$ is obtained
\cite{Brandenberger:2007ae, Vilenkin:1981zs, Gott:1984ef}.

We have considered the static configurations.
The extension to the stationary case where all the strings move with 
a constant velocity can be obtained by the boost transformation from
the static coordinate.
Because all the components of the Riemann tensor vanish in the
static coordinate,
they do in the stationary coordinate.
Therefore,
the geometry is also flat for the stationary configurations.

\section{Conclusion}
We showed,
without invoking the perturbative expansion, 
that the geometry around the stationary strings with junctions is flat,
which is a non-linear generalization of \cite{Brandenberger:2007ae}.

\bigskip
\bigskip

\noindent {\bf Acknowledgments:} 
The author would like to thank Masahiro Kawasaki for useful comments.


\begin{thebibliography}{100}
\bibitem{Dvali:2003zj}
  G.~Dvali and A.~Vilenkin,
  JCAP {\bf 0403}, 010 (2004)
  [arXiv:hep-th/0312007].

\bibitem{Copeland:2003bj}
  E.~J.~Copeland, R.~C.~Myers and J.~Polchinski,
  JHEP {\bf 0406}, 013 (2004)
  [arXiv:hep-th/0312067].

\bibitem{Leblond:2004uc}
  L.~Leblond and S.~H.~H.~Tye,
  JHEP {\bf 0403}, 055 (2004)
  [arXiv:hep-th/0402072].

\bibitem{Firouzjahi:2006vp}
  H.~Firouzjahi, L.~Leblond and S.~H.~Henry Tye,
  JHEP {\bf 0605}, 047 (2006)
  [arXiv:hep-th/0603161].

\bibitem{Dasgupta:2007ds}
  K.~Dasgupta, H.~Firouzjahi and R.~Gwyn,
  JHEP {\bf 0704}, 093 (2007)
  [arXiv:hep-th/0702193].

\bibitem{Jackson:2004zg}
  M.~G.~Jackson, N.~T.~Jones and J.~Polchinski,
  JHEP {\bf 0510}, 013 (2005)
  [arXiv:hep-th/0405229].

\bibitem{Copeland:2007nv}
  E.~J.~Copeland, H.~Firouzjahi, T.~W.~B.~Kibble and D.~A.~Steer,
  Phys.\ Rev.\  D {\bf 77}, 063521 (2008)
  [arXiv:0712.0808 [hep-th]].

\bibitem{Bevis:2008hg}
  N.~Bevis and P.~M.~Saffin,
  arXiv:0804.0200 [hep-th].

\bibitem{Rajantie:2007hp}
  A.~Rajantie, M.~Sakellariadou and H.~Stoica,
  JCAP {\bf 0711}, 021 (2007)
  [arXiv:0706.3662 [hep-th]].

\bibitem{Shlaer:2005ry}
  B.~Shlaer and M.~Wyman,
  Phys.\ Rev.\  D {\bf 72}, 123504 (2005)
  [arXiv:hep-th/0509177].

\bibitem{Brandenberger:2007ae}
  R.~Brandenberger, H.~Firouzjahi and J.~Karouby,
  Phys.\ Rev.\  D {\bf 77}, 083502 (2008)
  [arXiv:0710.1636 [hep-th]].

\bibitem{Vilenkin:1981zs}
  A.~Vilenkin,
  Phys.\ Rev.\  D {\bf 23}, 852 (1981).
  
\bibitem{Gott:1984ef}
  J.~R.~I.~Gott,
  Astrophys.\ J.\  {\bf 288}, 422 (1985).

\end{thebibliography}
\end{document}